\documentclass[12pt]{amsart}
\usepackage{amssymb}
\usepackage[polish,english]{babel}
\usepackage{lmodern}
\usepackage[utf8]{inputenc}
\usepackage[T1]{fontenc}
\usepackage{amsmath,amscd}
\usepackage[normalem]{ulem} 
\usepackage{amsaddr}

\usepackage{xcolor}
\usepackage{graphicx}
\usepackage{multicol}

\usepackage[section]{placeins}

\usepackage[hidelinks]{hyperref}
\newcommand{\doi}[1]{\\ \href{https://doi.org/#1}{\texttt{doi:#1}}}

\usepackage{url}

\usepackage{underscore}

\newcounter{question}

\providecommand{\code}[1]{\texttt{#1}}

\numberwithin{equation}{section}

\begin{document}

\author[P.\ Pilarczyk]{Pawe\l{} Pilarczyk}
\address{%P.\ Pilarczyk is with
Faculty of Applied Physics and  Mathematics \& Digital Technologies Center,
Gda\'{n}sk University of Technology,
%ul.\ Narutowicza 11/12,
80-233
Gda\'{n}sk, Poland.
%\\
Orcid: \href{https://orcid.org/0000-0003-0597-697X}{0000-0003-0597-697X}
}

\author[G.\ Graff]{Grzegorz Graff}
\address{%G. Graff is with 
Faculty of Applied Physics and Mathematics \& BioTechMed Center,
Gda\'{n}sk~University of Technology,
%ul.\ Narutowicza 11/12,
80-233
Gda\'{n}sk, Poland.
\\
(Corresponding author.)
%\\
Orcid: \href{https://orcid.org/0000-0001-5670-5729}{0000-0001-5670-5729}
}

\author[J.\ M.\ Amig\'{o}]{Jos\'{e} M. Amig\'{o}}
\address{
Centro de Investigaci\'{o}n Operativa (CIO), Universidad Miguel Hern\'{a}ndez, 03202~Elche, Spain.
%\\
Orcid: \href{https://orcid.org/0000-0002-1642-1171}{0000-0002-1642-1171}
}

\author[K.\ Tessmer]{Katarzyna Tessmer}
\address{%K.\ Tessmer is with
Faculty of Applied Physics and  Mathematics,
Gda\'{n}sk University of Technology,
%ul.\ Narutowicza 11/12,
80-233
Gda\'{n}sk, Poland.
%\\
Orcid: \href{https://orcid.org/0000-0003-4487-8037}{0000-0003-4487-8037}
}

\author[K.\ Narkiewicz]{Krzysztof Narkiewicz}
\address{%K.\ Narkiewicz is with
Department of Hypertension and Diabetology, Medical University of Gda\'{n}sk,  80-210~Gda\'nsk, Poland.
%\\
Orcid: \href{https://orcid.org/0000-0002-9181-9884}{0000-0002-9181-9884}
}

\author[B.\ Graff]{Beata Graff}
\address{%B. Graff is with
Department of Hypertension and Diabetology, Medical University of Gda\'{n}sk,  80-210~Gda\'nsk, Poland.
%\\
Orcid: \href{https://orcid.org/0000-0002-9181-9884}{0000-0002-9181-9884}
}

\title[Differentiating OSA patients by HR--BP coupling]{Differentiating patients with obstructive sleep apnea from healthy controls based on heart rate -- blood pressure coupling quantified by entropy-based indices}
\date{September 12, 2023}

\maketitle

%%%%%%%%%%%%%%%%%%%%%%%%%%%%%%%%%%%%%%%%%%%%%%%%%%%%%

\begin{abstract}
We introduce an entropy-based classification method for pairs of sequences (ECPS) for quantifying mutual dependencies in heart rate and beat-to-beat blood pressure recordings. The purpose of the method is to build a classifier for data in which each item consists of the two intertwined data series taken for each subject. The method is based on ordinal patterns, and uses entropy-like indices. Machine learning is used to select a subset of indices most suitable for our classification problem in order to build an optimal yet simple model for distinguishing between patients suffering from obstructive sleep apnea and a control group.
\end{abstract}

%%%%%%%%%%%%%%%%%%%%%%%%%%%%%%%%%%%%%%%%%%%%%%%%%%%%%

\bigskip

\noindent
\textbf{Heart rate (HR) and blood pressure (BP) are two important physiological variables that are tightly interrelated. Monitoring their mutual influence can provide important information about a person's cardiovascular health. In many cases, the altered effect of HR on BP or vice versa is a consequence of a serious disease. Therefore, indices that reflect mutual relation between these two quantities may support healthcare professionals in the development of more effective diagnostic and treatment approaches.
In particular, indices based on ordinal patterns and entropy can serve the purpose of quantitatively measuring mutual relation between two time series, as well as the complexity of each of them. Using these measures to quantify mutual dependencies between HR and BP sequences, we introduce a new method that involves machine learning techniques for differentiating between healthy subjects and persons suffering from obstructive sleep apnea.}

%%%%%%%%%%%%%%%%%%%%%%%%%%%%%%%%%%%%%%%%%%%%%%%%%%%%%

\section{Introduction}
\label{sec:intro}

%%%%%%%%%%%%%%%%%%%%%%%%%%%%%%%%%%%%%%%%%%%%%%%%%%%%%

In recent years, the interest in investigating the complex regulatory mechanisms of the human organism has notably increased. One of the most important directions of research in this area concerns studying interaction between various components of the cardiovascular system. This has created the need for more effective mathematical tools that would be useful in this kind of analysis.

The research goal of this paper is to develop an entropy-based classification method for the analysis of pairs of sequences (ECPS for short): heart rate (HR) and beat-to-beat blood pressure (BP), in order to differentiate between the group of healthy subjects and patients suffering from obstructive sleep apnea (OSA).

The problem of mutual relations between HR and BP is of growing interest in the community of exact sciences, and medical sciences as well. One of the main challenges in this kind of analysis is, according to \cite{Miller-rev}, \textit{extracting meaningful parameters usable for diagnostics and risk stratification}, and we aim at addressing this point with respect to the patients with OSA. 

OSA is a chronic condition characterized by pauses in breathing and/or decreases in airflow caused by repeated total or partial obstruction to the airway occurring during sleep. OSA is highly prevalent especially in patients with cardiovascular diseases. However, in clinical practice many times it is not recognized and not treated \cite{Yegh-OSA}.
Although the apnea-hypopnea index (AHI), calculated on the basis of a polysomnographic recording (PSG), remains the main tool for the diagnosis and planning of a treatment strategy in OSA patients, more detailed personal approach (``Precision sleep medicine'') is currently being advocated~\cite{Lim-Sleep}.

There are several studies of OSA in which classical linear measures are applied~\cite{Apnea-rev}, nonlinear methods have also been used in the last years on a wider scale~\cite{Varon-rev}.

It was shown that OSA leads to alterations in baroreflex which is one of the most important homeostatic mechanisms maintaining blood pressure. Moreover, previous findings suggested that baroreflex sensitivity is reduced before the onset of cardiovascular complications in OSA patients~\cite{Lombardi}.

We propose a comprehensive collection of indices based on ordinal patterns and entropy (OPE indices for short) that are known to be successful in the evaluation of mutual relations between sets of data.
We describe these indices in detail in Appendix~\ref{sec:indices}.

\begin{figure}[htbp]
\centering
\includegraphics[width=\textwidth]{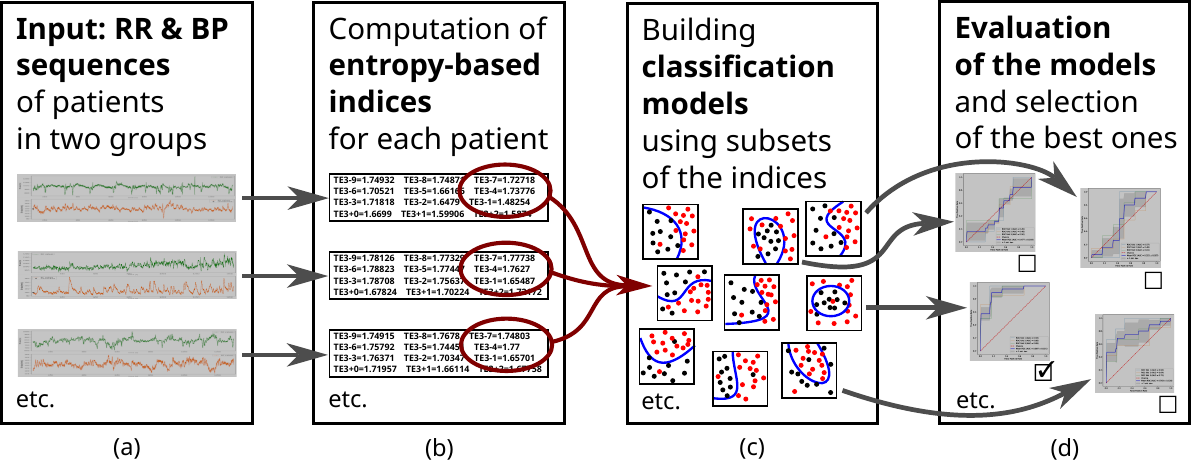}
\caption{\label{fig:overview}%
Overview of the ECPS method for constructing an optimal classifier of patients: (a) The input to the method consists of pairs of time series (HR and BP sequences) defined for each patient; the patients are split into two groups (CON and OSA). (b) For each patient separately, a collection of entropy-based indices are computed (see Appendix~\ref{sec:indices}). (c) Several classification SVM models are built using triplets of indices. (d) All the constructed models are evaluated using ROC curves, and the best ones are selected as optimal classifiers found using the ECPS method.}
\end{figure}

In real applications, many of the OPE indices may be correlated with each other. Moreover, the variability of some others may not be relevant to the actual differences between the data classes upon consideration. Therefore, in our ECPS method, we propose to use machine learning in order to select a small number of the most relevant indices, and to construct a simple nonlinear classifier that provides optimal results. In this way we develop a model that would distinguish patients that suffer from OSA from healthy subjects, based solely on the measurement of their HR and BP taken during an outpatient exam conducted during the day.
An overview of our approach is illustrated in Figure \ref{fig:overview}.

In \cite{Kurths-osa}, symbolic dynamics methods for the HR and BP sequences were successfully applied to study the cardiovascular regulation during different sleep stages in OSA syndrome.
Our approach, also based on the analysis of ordinal patterns, follows this promising line of research.
However, the novelty of our approach is threefold. 

First, we analyze the HR and BP sequences taken during wakefulness, that is, at the time when the sleep
apnea problem does not appear; however, long-term cardiovascular consequences of that might be detected.

 Second, we compute a large number of entropy-based indices. Last but not least, we use machine learning to select the most relevant indices which yield an optimal classifier.

Finally, we demonstrate the advantage of the selected OPE indices over models built upon traditional measures of HR and BP variability that are typically used in medical practice.

%%%%%%%%%%%%%%%%%%%%%%%%%%%%%%%%%%%%%%%%%%%%%%%%%%%%%

\FloatBarrier
\section{Materials and Methods}
\label{sec:appl}

%%%%%%%%%%%%%%%%%%%%%%%%%%%%%%%%%%%%%%%%%%%%%%%%%%%%%

\FloatBarrier
\subsection{Patients}
\label{sec:patients}

There were 19 patients diagnosed with obstructive sleep apnea (OSA) and  19 healthy volunteers (CON) included in the study. The participants were recruited in 2013--2018 from the outpatient hypertension clinic of the University Clinical Centre in Gda\'{n}sk and with the help of local advertisements. The study complied with the Declaration of Helsinki; a written informed consent was obtained from each studied person.
All of the subjects were males. There were no significant differences with respect to age between OSA and CON groups ($48.84\pm 6.48$ vs $51.11\pm 8.80$, respectively). Except for one OSA person, the patients were treated for hypertension but they did not differ significantly according to office blood pressure compared to the subjects in the CON group.

%%%%%%%%%%%%%%%%%%%%%%%%%%%%%%%%%%%%%%%%%%%%%%%%%%%%%

\FloatBarrier
\subsection{Obtaining raw RR and BP sequences}
\label{sec:RRandBP}

For each study participant, a 20-minute recording of ECG (with the PowerLab system and Lab Chart software, sampling rate 1000 Hz) and of non-invasive beat-to-beat blood pressure (with FINOMETER device) were taken in the supine position (except for one patient in the OSA group for whom a recording of 18 minutes is available only due to technical reasons). All the subjects were asked to relax but not to fall asleep.
We denote the two time series obtained for each patient $p$ as follows:
\begin{itemize}
\item $\{r_{p,i}\}_{i=0}^{k_p}$ -- a sequence of RR interval lengths, further called ``the RR sequence'' for short,
\item $\{b_{p,i}\}_{i=0}^{k_p}$ -- a sequence of systolic blood pressure (BP) readouts, further called ``the BP sequence'' for short.
\end{itemize}
The length of an RR interval is the distance in time between consecutive R peaks in the ECG signal recording. The associated \emph{time of its occurrence} is defined as the moment of occurrence of the second R in the ECG signal (the right end-point of the RR interval), and is denoted as $\text{time}(r_{p,i})$. The time of occurrence of each element in the BP sequence is denoted as $\text{time}(b_{p,i})$.

\begin{figure}
\flushright
\includegraphics[width=\textwidth]{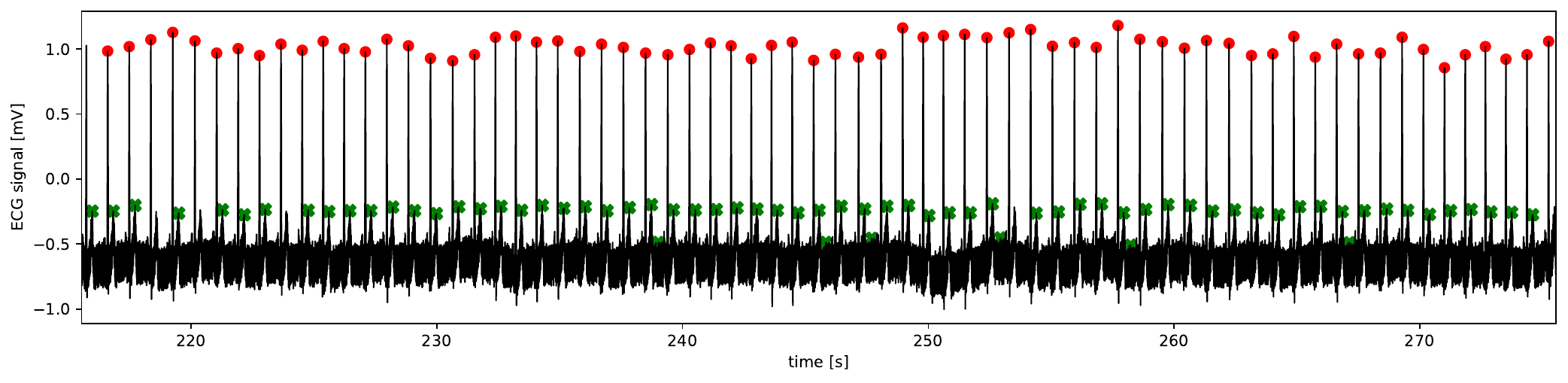}
\includegraphics[width=0.994\textwidth]{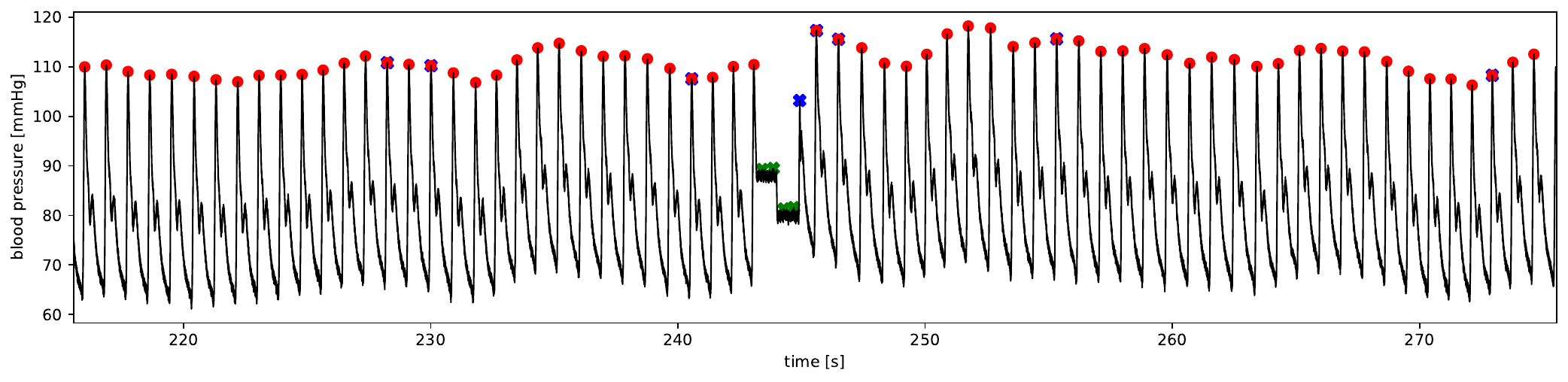}
\caption{\label{fig:signals}%
Samples of ECG (top figure) and blood pressure (bottom figure) signals with detected peaks marked with red dots. The green and blue crosses indicate other local maxima found in the signal that were excluded from the list of R's and systolic blood pressure readouts. An artifact caused by the blood pressure measuring device calibration procedure can be seen in the middle of the bottom figure.}
\end{figure}

Raw RR and BP sequences were automatically generated from raw ECG and blood pressure signals using heuristic algorithms; see Figure~\ref{fig:signals} for a real sample.

%%%%%%%%%%%%%%%%%%%%%%%%%%%%%%%%%%%%%%%%%%%%%%%%%%%%%

\FloatBarrier
\subsection{Cleaning and matching the RR and BP sequences}
\label{sec:cleaning}

The automatically detected R and BP peaks were visually inspected and manually corrected where necessary. Outliers in both sequences were marked as \emph{artifacts}. These sequences may also contain other points marked as artifacts or there might be some missing data points, which is due to signal disturbances or other kinds of problems, such as ectopic heartbeats (some detected automatically, some marked manually) in the RR sequence, or short periods of calibration of the blood pressure measuring device (detected automatically; shown in Figure~\ref{fig:signals}).

\begin{figure}
\centering
\includegraphics[width=\textwidth]{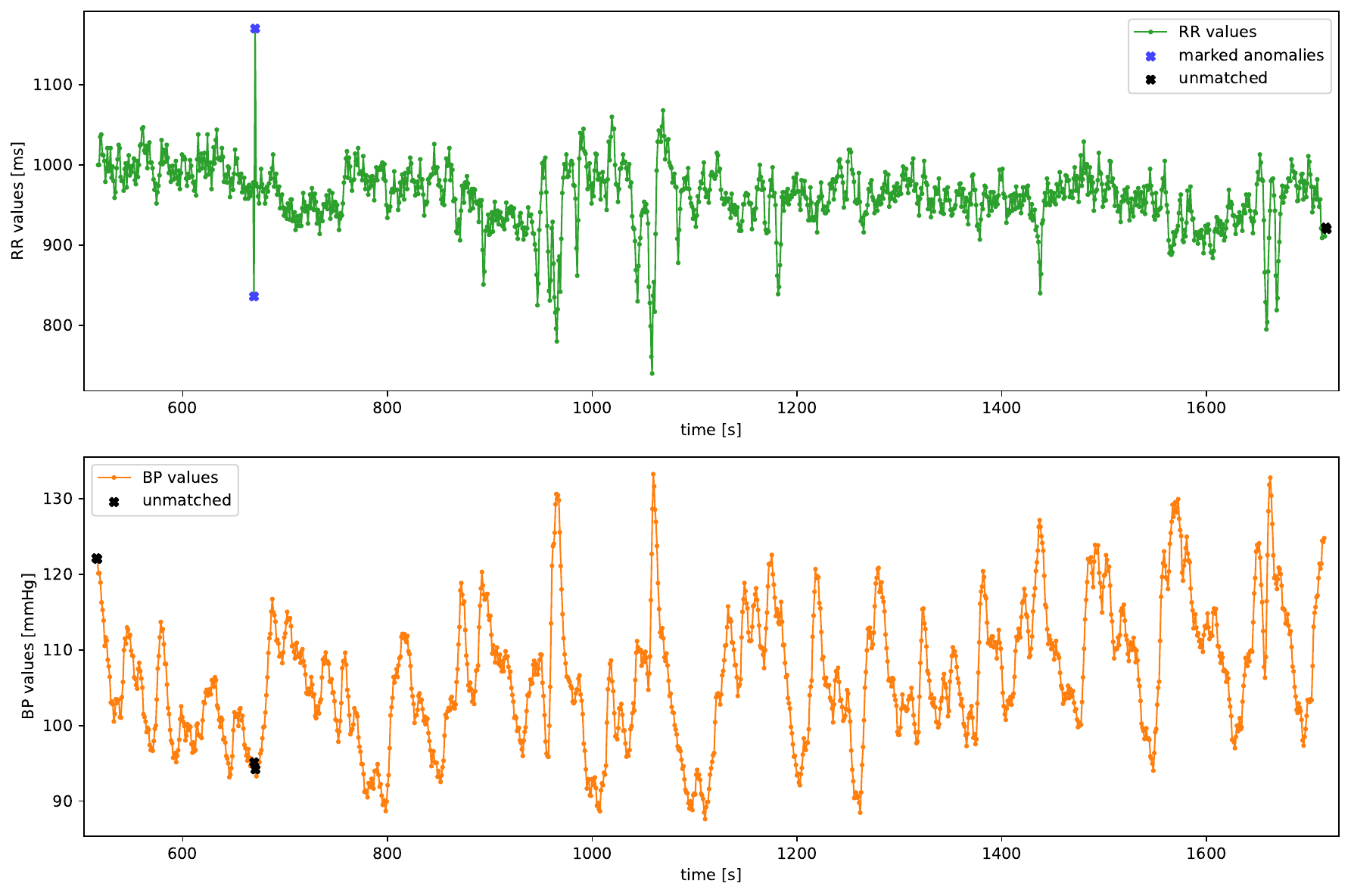}
\caption{\label{fig:matching}%
Matching an RR sequence with a BP sequence.
Two ectopic beats in the RR sequence (at time $670$~s) were marked as artifacts (blue crosses), which left the corresponding BP values unmatched (marked with black crosses).
There are also a few unmatched BP values at the beginning of the sequence, preceding the first RR detected, and some unmatched RR values at the end of the sequence; this kind of lack of matching often happens at the beginning or at the end of the recording and is normal.}
\end{figure}

These kinds of problems are unavoidable in the real clinical data. Therefore, in order to match the sequences in a reliable way for cause-and-effect analysis, their subsequences $\{r_{p,\kappa(i)}\}$ and $\{b_{p,\theta(i)}\}$ were selected in such a way that the chosen RR and BP readouts were not artifacts, and were interleaving in time: each RR interval occurrence time must be followed by a BP occurrence, which must in turn be followed by another RR interval occurrence, and so on. Specifically,
\[
\text{time}(r_{p,\kappa(i)}) <
\text{time}(b_{p,\theta(i)}) <
\text{time}(r_{p,\kappa(i+1)}) \qquad \text{for all $i$.}
\]
All the elements of the RR and BP sequences that were not included in these subsequences were marked as \emph{artifacts} by the software. See Figure \ref{fig:matching} for an example.

We use the two subsequences for further computations. Whenever a tuple of elements of any of the sequences is encountered whose indices are not consecutive (that is, there is some artifact between them), such a tuple is not taken into consideration.

%%%%%%%%%%%%%%%%%%%%%%%%%%%%%%%%%%%%%%%%%%%%%%%%%%%%%

\FloatBarrier
\subsection{Ensuring proper definition of ordinal patterns.}
\label{sec:bijection}

It is required in \eqref{ord patt} that each subsequence that defines an ordinal pattern must consist of pairwise different terms. However, we cannot expect that real time series that comes from measurements made with finite resolution would satisfy this assumption. One solution to this problem might be to put terms with lower indices first, but then certain patterns would be favored, which we would like to avoid. Therefore, we propose the following procedure (cf.\ \cite{Bandt2002} and the discussion about equal values in ordinal patterns in the recent paper \cite{Zanin 2021}).

Before computing ordinal patterns from time series, we add pairwise different small pseudo-random numbers to all the elements in all the sequences.   By ``small'' we mean numbers of 3 orders of magnitude smaller than the granularity of the specific sequence. The ``granularity'' is the smallest positive distance between any two terms in the sequence, not necessarily consecutive. For example, if we are given a sequence of integers then we add pseudo-random numbers from the open interval $(0,0.001)$.

Justification for this procedure is the following. In case of equal results of measurements, it may turn out that increasing the precision would reveal that one of the numbers is actually greater. However, we do not have this information, so we make a ``random choice'' in order to compensate for the indeterminacy.

%%%%%%%%%%%%%%%%%%%%%%%%%%%%%%%%%%%%%%%%%%%%%%%%%%%%%

\FloatBarrier
\subsection{Computation of the entropy-based indices}
\label{sec:indcomp}

We chose two different lengths $L$ of ordinal patterns obtained from the RR and BP series: $3$ and $4$, and the shift $\Lambda$ varying from $0$ to $9$. More precisely, we considered the following collection of quantities; see Appendix~\ref{sec:indices} for the definitions and explanations of the entropy-based indices listed here.

\begin{itemize}
\item permutation entropy for both RR and BP series, as defined by Equation~\eqref{PermEntr}: \\
\code{RR_PE3}, \code{RR_PE4}, \code{BP_PE3} and \code{BP_PE4} (note that all permutations are bijective);
\item statistical complexity for both RR and BP series, as defined by Equation~\eqref{SC5}: \\
\code{RR_SC3}, \code{RR_SC4}, \code{BP_SC3} and \code{BP_SC4};
\item transcript entropy between the RR and BP series shifted by $\Lambda \in \{0, 1, \ldots, 9\}$, as defined by Equation \eqref{cross-transcript ent}: \\
\code{TE3+0}, \ldots, \code{TE3+9} and \code{TE4+0}, \ldots, \code{TE4+9}, \\
or between the BP and RR series shifted by $\Lambda \in \{1, 2, \ldots, 9\}$: \\
\code{TE3-1}, \ldots, \code{TE3-9} and \code{TE4-1}, \ldots, \code{TE4-9};
\item self-transcript entropy of the RR and BP series with the shift $\Lambda \in \{1, 2, \ldots, 9\}$, as defined by Equation \eqref{self-transcript ent}: \\
\code{RR_STE3+1}, \ldots, \code{RR_STE3+9}, and \code{RR_STE4+1}, \ldots, \code{RR_STE4+9}, \\
as well as \code{BP_STE3+1}, \ldots, \code{BP_STE3+9} and \code{BP_STE4+1}, \ldots, \code{BP_STE4+9};
\item mutual information between RR and BP series shifted by $\Lambda \in \{-9, \ldots, 9\}$, as defined by Equation \eqref{Perm Mutual Info}: \\
\code{MI3-9}, \ldots, \code{MI3+9} and \code{MI4-9}, \ldots, \code{MI4+9};
\item transcript mutual information from RR to BP with the shift $\Lambda \in \{1, 2, \ldots, 9\}$, as defined by Equation \eqref{TranscriptMI}; see also \cite[formula (19)]{Amigo2015}: \\
\code{TMI3+1}, \ldots, \code{TMI3+9} and \code{TMI4+1}, \ldots, \code{TMI4+9}, \\
and also from BP to RR with the same range of shifts: \\
\code{TMI3-1}, \ldots, \code{TMI3-9} and \code{TMI4-1}, \ldots, \code{TMI4-9}.
\end{itemize}

With the above selection of entropy-based indices, we expect to capture most characteristics of the signals to be analyzed below. Indeed, entropy is a measure of randomness, achieving its maximum value for equiprobable ordinal patterns and transcripts, e.g., white noise. On the other hand, statistical complexity focuses rather on the ``structure'' of the signal, taking the largest values for signals with intermediate normalized permutation entropies and vanishing for both white noise and zero-entropy signals \cite{Martinez2022}. As for the mutual information, it is a nonlinear measure of dependence between two random variables~\cite{Cover2006}, which is symmetric in the conventional case and asymmetric if transcripts are used; as said in Section~\ref{sec:mutual}, transcript mutual information is a causality indicator. In addition, all these indices
are model-free, although they depend on one or two parameters: $L$ (the
length of the ordinal patterns) and $\Lambda $ (coupling delay). The values
of $L$ and $\Lambda $ were chosen based on our experience \cite{Amigo2015,Amigo2016,Amigo2019,Graff2013} and, as shown by the results, they
turned out to be satisfactory for our purposes.

%%%%%%%%%%%%%%%%%%%%%%%%%%%%%%%%%%%%%%%%%%%%%%%%%%%%%

\FloatBarrier
\subsection{Computation of the classical HRV and BPV indices}
\label{sec:classical}

For comparison, several classical heart rate variability (HRV) parameters were also computed for the RR series using the ``hrv-analysis'' Python module.\footnote{Heart Rate Variability analysis, \url{https://github.com/Aura-healthcare/hrv-analysis/}.}
% https://aura-healthcare.github.io/hrv-analysis/hrvanalysis.html
They were computed for the RR series after exclusion of artifacts (such as outliers or ectopic beats, as explained in Section~\ref{sec:cleaning}). We analyzed the following parameters:
\begin{itemize}
\item time domain features: \code{mean_nni}, \code{sdnn}, \code{sdsd}, \code{pnni_50}, \code{pnni_20}, \code{rmssd}, \code{median_nni}, \code{range_nni}, \code{cvsd}, \code{cvnni}, \code{mean_hr}, \code{max_hr}, \code{min_hr}, \code{std_hr};
\item geometrical features: \code{triangular_index}, \code{tinn};
\item CSI CVI features: \code{csi}, \code{cvi}, \code{Modified_csi};
\item Poincar\'{e} plot features: \code{sd1}, \code{sd2}, \code{ratio_sd2_sd1};
\item frequency domain features: \code{lf}, \code{hf}, \code{lf_hf_ratio}, \code{lfnu}, \code{hfnu}, \code{total_power}, \code{vlf}.
\end{itemize}
Note that we excluded sample entropy from the list of the classical HRV parameters, because this already is an entropy-based index, and our intention was to compare our entropy-based indices with indices of different type.

We also computed the following quantities for determining the degree of blood pressure variability (BPV):
\begin{itemize}
\item time domain features: \code{mean}, \code{sd};
\item frequency domain features: \code{hf}, \code{lf}, \code{vlf};
\end{itemize}
see \cite[Figure 2 and Table 2]{Parati2023} for justification of our selection of the features in addition to the most obvious choice of the mean and standard deviation. Similarly to the HRV indices, we did not include sample entropy in our comparison. The results of the computation of all these indices are available in~\cite{www}.

%%%%%%%%%%%%%%%%%%%%%%%%%%%%%%%%%%%%%%%%%%%%%%%%%%%%%

\FloatBarrier
\subsection{Choosing an optimal set of indices}
\label{sec:svmTriplets}

The OPE indices listed in Section~\ref{sec:indcomp} give rise to $156$ variables that can be taken as a basis for statistical analysis and machine learning. In our ECPS method, we are specifically interested in building an effective classifier.

Since it happens that variables irrelevant to the classification may degrade the quality of the classifier, we propose to choose a small subset of the indices that would yield an optimal classifier. Since it is not obvious which lengths $L$, which coupling delays $\Lambda$, and which specific indices are optimal for a specific application, we propose to test all the possible triplets of indices, and choose the triplet that provides optimal results. These results can be assessed using a chosen method for the assessment of the classification model of the type that we use.

Taking this into consideration, we used receiver operating characteristic curves (ROC curves) and we computed average area under the curve (AUC) for the ROC curves obtained in 5 repetitions of a 3-fold cross-validation test for the non-linear support vector machine (SVM) binary classifier with the radial basis function (RBF) kernel (see e.g. \cite{HanKamberPei}) for all the possible triplets of OPE indices.
Although it may seem a computationally demanding task, in fact all these computations were completed by a Python script on a modern PC within less than $30$ minutes. After the model has been constructed, applying it to new data (e.g., diagnosing another patient) is very fast.

More specifically, we generated the collection $C$ of all the $620{,}620$ subsets of cardinality $3$ of the set of all the $156$ features (such as \texttt{MI4+3}) computed for the ordinal patterns, listed in Section~\ref{sec:indcomp}. For each set $c \in C$ we conducted the following test. We split the set of subjects into $3$ subsets of approximately equal size with approximately the same proportion of subjects in both groups (which is known under the term ``with stratification''). We then built a classifier using the standard nonlinear SVM model with the RBF kernel on two of these three sets, and tested this model on the third set. The quality of the classifier can be assessed visually on the basis of the ROC curves, and numerically by computing the area under the curve (AUC). The higher the AUC value, the better the classifier: the value of about $0.5$ indicates a classifier that is as good as random choice, while the value of $1.0$ indicates the perfect classifier that makes no mistakes. We computed the mean of the three AUCs. We repeated this procedure of training the SVM model and testing it for the three possibilities of choosing the sets for building the model. This procedure of splitting the set of subjects, building the classifier, and testing it was repeated $5$ times in order to ensure that the good or bad result is not due to a more or less lucky subdivision of the set of subjects into the three subsets. In this way, for each triplet of the indices under consideration, we computed the mean AUC, the standard deviation of all the $15$ AUCs obtained in the $5$ repetitions, as well as the minimum and the maximum values of the five mean AUCs. The computed values were gathered in a table, which we then sorted by mean AUC.

%%%%%%%%%%%%%%%%%%%%%%%%%%%%%%%%%%%%%%%%%%%%%%%%%%%%%

\section{Results}
\label{sec:results}

Recall that whenever any of the ordinal patterns under analysis contained an artifact in the RR or BP series, its appearance was not counted. The percentage of this kind of situations was recorded. 
In our dataset, there were no ordinal patterns omitted in the sequences for $21$ subjects. The number of omitted patterns was below $1\%$ in the data of further $11$ subjects, below $2\%$ for the next $5$ subjects, and reached the maximum of $2.6\%$ for one subject.
In total, less than $0.305\%$ of all the patterns were omitted in all the analyses. The results of the computation of all these indices are available in~\cite{www}.

Before proceeding with machine learning, we applied the U Mann-Whitney test to verify whether any single index could be used alone to differentiate between the two groups of patients (with $p$-value below $0.05$). The result was positive for two classical HRV indices: \code{HRV\_mean\_nni} ($p$-value $0.0308$) and \code{HRV\_median\_nni} ($p$-value $0.0329$), and four OPE indices: \code{TMI4-9}, \code{TMI4+5}, \code{TMI4+6}, \code{TMI4+7} ($p$-values between $0.0102$ and $0.0307$).

We also computed Pearson linear correlations (with $p$-values) between all the OPE and classical HRV and BPV indices on the set of all the subjects together and also on the CON and OSA groups. The results are gathered in CSV files and in illustrations available in~\cite{www}. We found several strong correlations between some HRV indices, mainly due to the way they were defined, and other correlations that were well known already. The classical BPV indices did not exhibit such strong correlations with any of the classical HRV or BPV indices. In the case of the OPE indices that we computed, however, we found a considerable number of very strong correlations, in some cases even above $0.98$ in absolute value. Finally, we verified correlations between the OPE indices and the classical HRV and BPV indices, but we did not find very strong correlations there.

The best triplets of the indices derived from ordinal patterns following the procedure described in Section~\ref{sec:svmTriplets} are listed in Table~\ref{tab:bestORD}, ordered by mean AUC. The table also contains the standard deviation of the AUCs computed in the five attempts, as well as the minimum and the maximum mean AUC encountered in the five attempts.

\begin{table}[htbp]
\centering
\begin{tabular}{cccl}
Mean AUC $\pm$ SD & Min AUC & Max AUC & the triplet of indices \\
\hline
$0.857 \pm 0.086$ & 0.800 & 0.897 & \code{RR\_STE3+6,MI3+4,MI4+9} \\
$0.853 \pm 0.070$ & 0.811 & 0.885 & \code{RR\_STE3+6,MI3+4,MI4+8} \\
$0.846 \pm 0.072$ & 0.800 & 0.875 & \code{RR\_STE3+6,MI3+5,MI4+8} \\
$0.846 \pm 0.076$ & 0.794 & 0.888 & \code{RR\_STE3+6,MI3+4,MI4+7} \\
$0.838 \pm 0.068$ & 0.790 & 0.879 & \code{MI3+4,RR\_STE4+5,MI4+8} \\
$0.833 \pm 0.104$ & 0.737 & 0.870 & \code{RR\_STE3+6,MI3+5,MI4+7} \\
$0.831 \pm 0.113$ & 0.741 & 0.876 & \code{RR\_STE3+6,MI4+7,MI4+8} \\
$0.831 \pm 0.088$ & 0.776 & 0.895 & \code{MI3+4,RR\_STE4+5,MI4+9} \\
$0.822 \pm 0.094$ & 0.779 & 0.838 & \code{RR\_STE3+6,MI3+5,MI4+9} \\
$0.819 \pm 0.078$ & 0.780 & 0.857 & \code{RR\_STE3+4,RR\_STE3+6,MI4+8} \\
\hline
\end{tabular}
\vskip 12pt
\caption{\label{tab:bestORD} The triplets of OPE indices that yield the best non-linear SVM models based on the OPE indices.}
\end{table}

In order to obtain a reference point to benchmark the performance of the OPE indices, an analogous computation was conducted using the $33$ classical HRV and BPV features described in Section~\ref{sec:classical} ($5{,}456$ triplets), and the best results are shown in Table~\ref{tab:bestHRV}.

\begin{table}[htbp]
\centering
\footnotesize
\begin{tabular}{cccl}
Mean AUC $\pm$ SD & Min AUC & Max AUC & the triplet of indices \\
\hline
$0.694 \pm 0.137$ & 0.614 & 0.763 & \code{HRV\_range\_nni,HRV\_min\_hr,HRV\_triangular\_index} \\
$0.691 \pm 0.161$ & 0.532 & 0.779 & \code{HRV\_min\_hr,HRV\_triangular\_index,BPV\_sd} \\
$0.689 \pm 0.120$ & 0.583 & 0.762 & \code{HRV\_pnni\_20,HRV\_range\_nni,HRV\_triangular\_index} \\
$0.669 \pm 0.144$ & 0.548 & 0.771 & \code{HRV\_median\_nni,HRV\_range\_nni,HRV\_triangular\_index} \\
$0.668 \pm 0.138$ & 0.599 & 0.730 & \code{HRV\_cvnni,HRV\_min\_hr,HRV\_Modified\_csi} \\
$0.667 \pm 0.107$ & 0.562 & 0.751 & \code{HRV\_pnni\_20,HRV\_range\_nni,HRV\_cvnni} \\
$0.665 \pm 0.100$ & 0.592 & 0.728 & \code{HRV\_pnni\_20,HRV\_cvnni,HRV\_lf} \\
$0.664 \pm 0.140$ & 0.572 & 0.746 & \code{HRV\_range\_nni,HRV\_mean\_hr,HRV\_triangular\_index} \\
$0.661 \pm 0.151$ & 0.566 & 0.755 & \code{HRV\_cvnni,HRV\_min\_hr,HRV\_lf} \\
$0.660 \pm 0.139$ & 0.544 & 0.755 & \code{HRV\_mean\_nni,HRV\_range\_nni,HRV\_triangular\_index} \\
\hline
\end{tabular}
\vskip 12pt
\caption{\label{tab:bestHRV} The triplets of indices that yield the best non-linear SVM models built upon the classical HRV and BPV indices.}
\end{table}

The results of the computation of all these triplets of indices are available in~\cite{www}. One CSV file contains all the OPE triplets, and another CSV file contains all the classical HRV and BPV triplets, together with the data indicated in Tables \ref{tab:bestORD} and~\ref{tab:bestHRV}, sorted by the mean AUC value in the descending order.

\begin{figure}[htbp]
\centering
\includegraphics[width=0.6\textwidth]{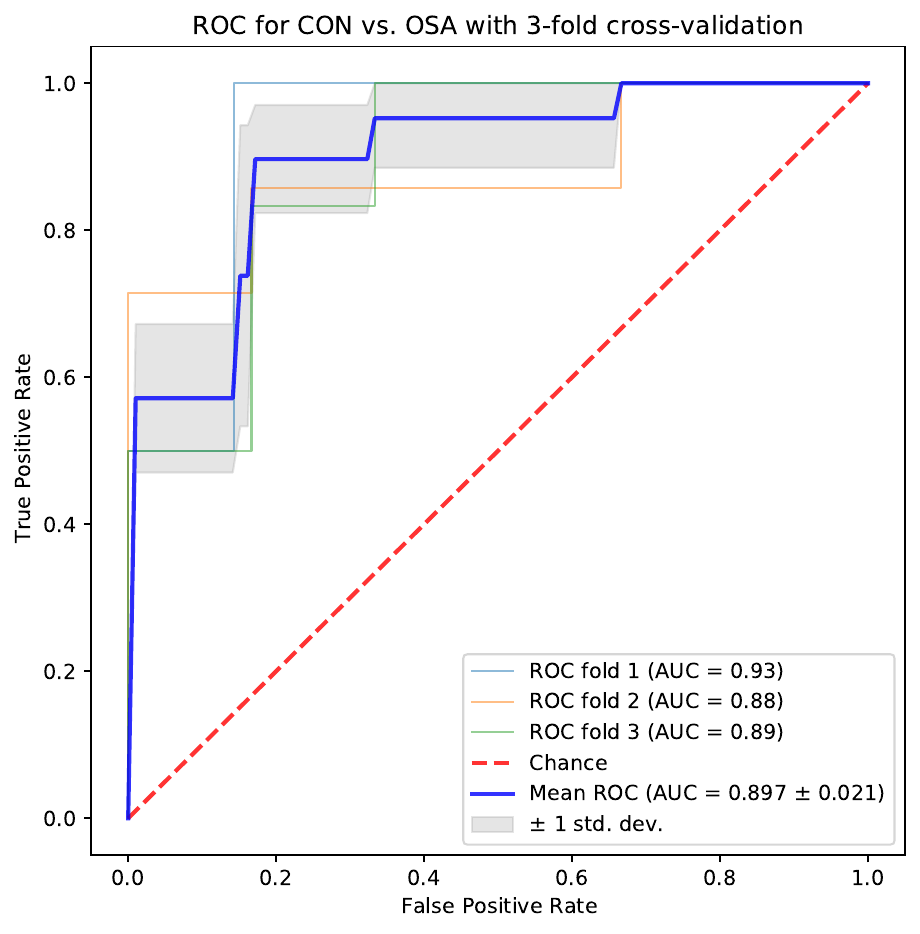}
\caption{\label{fig:bestROC_ord}%
Results for one of the nonlinear SVM classification models built upon the top-ranked triplet of OPE indices: \code{RR\_STE3+6}, \code{MI3+4}, \code{MI4+9}.}
\end{figure}

\begin{figure}[htbp]
\centering
\includegraphics[width=0.6\textwidth]{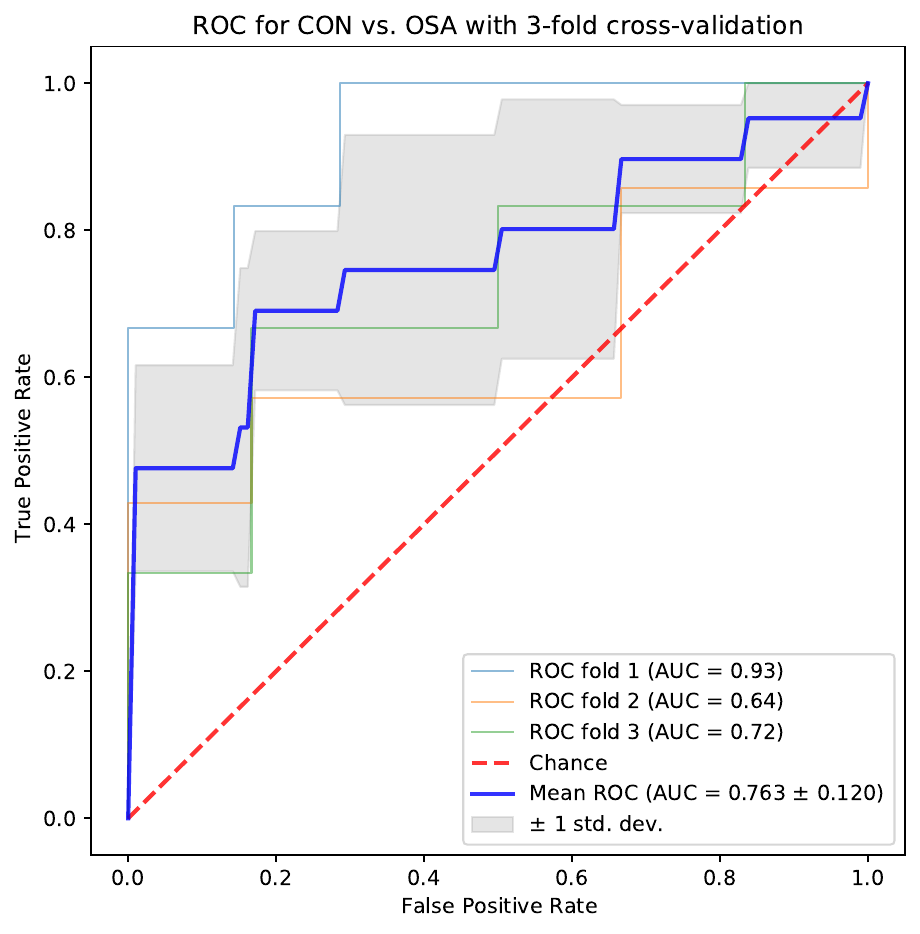}
\caption{\label{fig:bestROC_hrvNOe}%
Results for one of the nonlinear SVM classification models built upon the top-ranked triplet of HRV and BPV indices: \code{HRV\_range\_nni}, \code{HRV\_min\_hr}, \code{HRV\_triangular\_index}.}
\end{figure}

The ROC curves and AUCs computed for the best of the 5 cases of the 3-fold cross validation for the best triplets of indices are shown in Figure~\ref{fig:bestROC_ord} for the OPE indices, and in Figure \ref{fig:bestROC_hrvNOe} for the others.

For the purpose of measuring the overall performance of the specific indices that participated in all the tested triplets, we computed the average value of $1\%$ of the best mean AUCs in which each of the features was included. The best ten OPE indices turned out to be
\code{RR\_STE3+6},
\code{TMI4+1},
\code{RR\_STE4+5},
\code{MI4+9},
\code{MI4+8},
\code{RR\_STE3+3},
\code{BP\_STE3+3},
\code{TE3+1},
\code{MI4-5},
and \code{MI4+7}
(in this order), and the best ten classical HRV and BPV indices were:
\code{HRV\_pnni\_20},
\code{HRV\_triangular\_index},
\code{BPV\_sd},
\code{HRV\_range\_nni},
\code{HRV\_cvi},
\code{HRV\_Modified\_csi},
\code{BPV\_vlf},
\code{HRV\_min\_hr},
\code{HRV\_cvnni},
and \code{HRV\_lf},]
(in this order).
It is interesting to see that the indices successful in statistical tests are not present in these lists. Indeed, as far as the \code{TMI4}-based indices are concerned, their values are statistically different in the two groups, but the ranges overlap considerably, which apparently does not help in building a successful classifier.

\begin{figure}[htbp]
\centering
\includegraphics[width=0.7\textwidth]{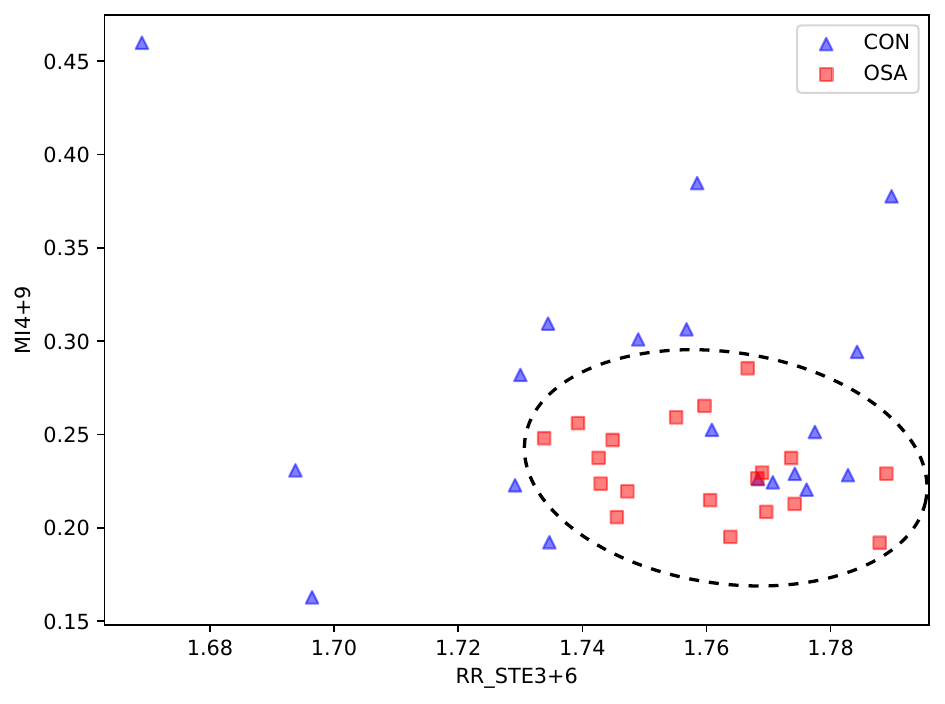}
\caption{\label{fig:plotORD1}%
Data points projected onto the plane spanned by two OPE indices \code{RR\_STE3+6} and \code{MI4+9} that participate in the best triplet found. Note the relatively good non-linear separation of the groups.}
\end{figure}

The quality of the indices found in the best triplets can be illustrated by the fact that one can see relatively good separation of the data groups already in a $2$-dimensional plot in two selected coordinates; see Figure~\ref{fig:plotORD1} for an example. We would like to remark that it is a common practice to apply PCA in order to visualize the separation of data groups in the plane of greatest variability of the variables. We made such an attempt using all the OPE indices together, but -- although we do not show the corresponding figures here -- we did not see any separation in the first few components. This proves that variability of some of the indices was in fact useless for distinguishing the two groups. Our machine learning approach helped us ignore such indices and extract the useful ones, so that we obtained satisfactory effect by simply selecting two of the optimal variables found.

%%%%%%%%%%%%%%%%%%%%%%%%%%%%%%%%%%%%%%%%%%%%%%%%%%%%%

\section{Disscussion}
\label{sec:disscussion}

Our results, including the application of the ECPS method (Section~\ref{sec:svmTriplets}) to the two groups of subjects, confirm the importance of conducting the analysis of mutual relation between heart rate variability (HRV) and blood pressure variability (BPV), particularly if one aims at the development of machine learning methods that could support one in diagnosing certain diseases. With the help of several indices based on ordinal patterns and entropy (OPE indices), we developed a satisfactory classifier capable of distinguishing patients that suffered from obstructive sleep apnea (OSA) from a control group (CON). A similarly built classifier that was based on the classical analysis of HRV and BPV sequences without delving into their mutual relationship (and without considering the entropy) achieved considerably worse results.

We would like to point out the fact that in order to measure the quality of the classifier in Section~\ref{sec:svmTriplets}, we did not only look at mean AUC, but also at the consistency between results obtained for different train--test splits of the data, reflected by their standard deviation that is shown in the tables. Indeed, as one can see in Tables \ref{tab:bestORD} and~\ref{tab:bestHRV},
some results with relatively high AUCs in fact corresponded to models whose reliability could be questioned, because in spite of high AUC for some subdivisions of the set of subjects, in some other cases the results were considerably worse. This is very well reflected in Figures \ref{fig:bestROC_ord} and~\ref{fig:bestROC_hrvNOe}, where one can see the wide gray belt and one poor classification result (AUC=$0.64$) for the standard HRV and BPV indices, as opposed to constantly valuable results achieved for the OPE indices.

Correlations between the OPE indices and the classical HRV and BPV indices might provide some hints for the interpretation of the actual physiological features that were reflected by the OPE indices. In particular, stronger correlations in general could be noticed in the CON group than in the OSA group, which might indicate impaired regulation between HR and BP in the latter patients.

At this point, we would like to make a comment on why we chose to investigate triplets of indices, and not pairs or quadruplets (or larger tuplets). In fact, we did conduct analogous computations for pairs and quadruplets of indices. However, the results for pairs were clearly worse than those for the triplets, while the computations for quadruplets yielded essentially as good results as for the triplets, but were considerably more costly in terms of computation time, because we had to run each individual machine learning test for each of the $23{,}738{,}715$ quadruplets, as opposed to the $620{,}620$ triplets. We suspect that the fact that the results were not significantly improved could be attributed to strong correlation (negative or positive) between many OPE indices, which one can easily see in the illustrations provided in~\cite{www}.

It might also be tempting to build a classification model based on all the indices together. Indeed, we conducted such an experiment. We used all the $156$ OPE indices together for one model, and all the $33$ classical HRV and BPV indices together for the other model. We tried both linear SVM and the nonlinear SVM with the RBF kernel. The results of 3-fold cross-validation in all the tested cases were discouraging -- the quality of the SVM classifier was essentially equivalent to the ``random choice'' classifier, as the resulting mean AUCs were rarely above $0.6$.
This shows why it is beneficial to choose a selection of indices in the ECPS method, as described in Section~\ref{sec:svmTriplets}.

We would like to emphasize the fact that the ECPS method may be enriched by adding other types of indices, and then using the wider selection for choosing an optimal subset.
Indeed, the interaction between time series that represent HR and BP has been observed and analyzed using a variety of methods in many recent studies, such as \cite{Faes-mechanism,Faes2011Comp,Faes2011Front,JavorkaRRBP2,Marwan-recurrence,PortaAMJPICP2011,JavorkaRRBP1,Wessel2020}; see also the review papers \cite{Miller-rev,Schulz-rev} and the references therein. For example, momentary information transfer was introduced in~\cite{PompeRunge2011} for multivariate analysis of coupling between two time series. In order to verify whether adding this index to our analysis could yield better results, we additionally computed this index with delays in $\{-9, \ldots, 9\}$, as defined by \cite[Formula (9)]{PompeRunge2011}; we shall further denote these indices as \code{MIT-9}, \ldots, \code{MIT+9}. Then we applied the analysis described in Section~\ref{sec:svmTriplets} to the resulting set of $175$ features (instead of the original $156$ ones). Some of the new indices indeed appeared in the tested triplets at relatively high positions, e.g., \code{MIT-7} in a triplet yielding AUC $0.835$ (position 6 in the ranking) and AUC $0.831$ (position 8), and \code{MIT-4} in a triplet yielding AUC $0.824$ (position 10). This indicates the usefulness of the \code{MIT} indices, although adding them did not improve the best triplet found.

Another question is whether our method is specific to the OSA patients, or maybe it might be applied to other pairs of time series, too. In order to verify this, we tested our method on an artificial set of $20$ RR and BP sequences generated by bidirectionally coupled logistic maps, as defined in \cite[Formula (25) and Figure 2(a)]{PompeRunge2011}. Already using basic statistical tests, a multitude of the OPE indices, and also all the \code{MIT} indices allowed to significantly differentiate between the group of these pairs of sequences and a set of $20$ other pairs of sequences generated by uncoupled logistic maps. This shows that with machine learning support, the ECPS method is capable of finding much more subtle differences in pairs of time series than those discussed in~\cite{PompeRunge2011}.

The quantitative definitions and qualitative meanings of the OPE indices were given in Appendix~\ref{sec:indices} and Section~\ref{sec:indcomp}, respectively. It is remarkable that the best four OPE indices in statistical tests (Section~\ref{sec:results}) were based on transcripts; namely, this was the transcript mutual information with different coupling delays. Similarly, among the best ten OPE indices that participated in all the tested triplets (Section \ref{sec:results}), six of them were also based on transcripts; namely, self-transcript entropy appeared four times, and transcript mutual information and transcript entropy appeared once. We conclude that transcripts have a noteworthy potential for discrimination.
Note that the SC indices did not participate in high ranked classification models, even though they are often taken into consideration in the literature.
It is also a noticeable fact that among all the classical HRV and BPV indices, the only statistically significant differences between the groups were exhibited by the mean and median length of time intervals between heartbeats.

The results illustrated in Figure \ref{fig:plotORD1} showed that, when considering simultaneously both indices \code{RR_STE3+6} and \code{MI4+9}, patients with OSA have lower values of \code{MI4+9} and the same or greater values of \code{RR_STE3+6}. Recall that \code{MI4+9}, the mutual information, is a symmetric measure of dependence. Its lower values indicate  weaker coupling that reflects impaired information flow between RR and BP series. 
The greater values of \code{RR_STE3+6} are related to higher predictability of the RR series of OSA patients, which may be interpreted as a loss of complexity in heart rate in comparison of healthy subjects, a typical syndrome in pathology or aging.
Let us remark that we observed the same regularities in case of 
data points projected onto the plane spanned by two other OPE
indices: \code{RR_STE3+6} and \code{MI3+4} that are present in the best triplet found. 

It might be tempting to create a ``profile'' of a healthy patient with respect to the indices that we computed, or to indicate ranges of values of concern with respect to OSA. However, the separation of the two groups of patients is not linear, as shown in Figure~\ref{fig:plotORD1}, which makes description of such a profile cumbersome. Moreover, the number of patients tested in our research is too small to make a reliable profile; a more involved research would be necessary. Nevertheless, based on our results one might make an attempt, for example, to claim that \code{RR\_STE3+6} values between $1.74$ and $1.79$ might be an indicator of possible OSA problems, provided that \code{MI4+9} is within $0.17$ and $0.27$ and \code{MI3+4} is between $0.03$ and $0.08$.

The main conclusion from the specific application of our ECPS method to the RR and BP time series is that the OPE methods might show altered cardiovascular regulation in awake OSA patients. 
Although there is a number of studies on HRV and BPV in OSA patients, they differ with the time of assessment (mostly night-time), parameters tested (usually HR only) and indices used for the parameters' variability analysis (typically standard, linear methods) \cite{Qin-OSA,Ucak}.

Non-linear parameters have been recently proven to have a great potential for showing specific features of HRV in a large group of OSA patients. With the increasing severity of the disease, lower complexity of heart rate during sleep \cite{Liang} was found in OSA subjects, and, what is more, also during wakefulness \cite{Qin-OSA-HRV}.

Studying the relation between heart rate and blood pressure in continuous recordings might be challenging. Most of the studies in OSA subjects in this area were based on the assessment of baroreflex sensitivity (BRS). The question remains, however, whether the standard BRS assessment is able to catch the whole range of autonomic disturbances in OSA patients \cite{Lombardi}.

Another approach has been proposed recently that allows the assessment of non-linear nature of cardiovascular regulation based on information transfer quantification  \cite{Faes2011Comp, Faes2011Front, JavorkaRRBP2, JavorkaRRBP1}. 
To our best knowledge, this is the first study which successfully combines the above kind of approach with the use of advanced symbolic methods and machine learning for the elucidation of changes in awake OSA patients.
However, the interpretation of physiological meaning of OPE indices is challenging and needs further investigation in a larger group of patients.  For example, in our study the best discrimination power was achieved by the MI and TMI indices for larger lags, which cannot be directly explained by  baroreflex or Frank-Starling mechanisms. We hypothesized that in group of OSA patients altered respiration has to be considered first looking for the explanation.

In our recent pilot study, patients with apnea episodes and desaturations characterized with the increased respiratory variability in short recordings during wakefulness which could impact cardiovascular coupling~\cite{GraffESGCO}. 
There is also an emerging area of growing interest considering various vascular or blood pressure response to sympathetic outflow (i.e., sympathetic transduction), depending for example on age, sex or obesity status, which might explain the delay of 6--9 heartbeats. It was also hypothesized that altered sympathetic transduction might be a compensatory mechanism in subjects with baroreflex dysfunction; in fact, it is known that baroreflex impairment is present in OSA patients~\cite{Young}.
Summarizing, further studies are needed to assess the actual mechanisms underlying our findings related to the role of long delays as well as the effectiveness of the ordinal pattern parameters in differentiating the two groups of subjects that we considered.

%%%%%%%%%%%%%%%%%%%%%%%%%%%%%%%%%%%%%%%%%%%%%%%%%%%%%

\section*{Data Availability}

The data that support the findings of this study are available from the corresponding authors upon reasonable request.

%%%%%%%%%%%%%%%%%%%%%%%%%%%%%%%%%%%%%%%%%%%%%%%%%%%%%

\section*{Acknowledgments}

This research was supported by the National Science Centre, Poland, within the following grants:
SHENG 2018/30/Q/ST1/00228 (G. Graff), OPUS 2021/41/B/ST1/00405 (P. Pilarczyk), and MAESTRO 2011/02/A/NZ5/00329 (B. Graff and K. Narkiewicz). J.M. Amig\'{o} was financially supported by Agencia Estatal de Investigaci\'{o}n,
Spain, grant PID2019-108654GB-I00/AEI/10.13039/501100011033 and by
Generalitat Valenciana, Spain, grant PROMETEO/2021/063. 
G. Graff and K. Tessmer were also supported by
Gda\'nsk University of Technology grant Neptunium Excellence Initiative - Research University, DEC -2/2021/IDUB/II.
% the acknowledgment below is required by TASK; see: https://docs.task.gda.pl/kdm/poradnik-uzytkownika/korzystanie-z-kdm/
Computations were carried out using the computers of Centre of Informatics Tricity Academic Supercomputer \& Network.

%%%%%%%%%%%%%%%%%%%%%%%%%%%%%%%%%%%%%%%%%%%%%%%%%%%%%

%%%%%%%%%%%%%%%%%%%%%%%%%%%%%%%%%%%%%%%%%%%%%%%%%%%%%

\appendix

\section{Mathematical tools: Ordinal patterns and entropy-based indices (OPE indices)}
\label{sec:indices}

Let $(\Omega ,\mathcal{B},\mu )$ be a probability space where, for the sake
of this paper, $\Omega $ is a compact manifold, $\mathcal{B}$ is the Borel
sigma-algebra of $\Omega $, and $\mu $ is a probability over the measurable
space $(\Omega ,\mathcal{B})$. If, furthermore, the map $f\colon \Omega
\rightarrow \Omega $ is $\mu $-invariant, i.e., $f$ is measurable and $\mu
(f^{-1}B)=\mu (B)$ for all $B\in \mathcal{B}$, then $(\Omega ,\mathcal{B}%
,\mu ,f)$ is called a (discrete-time, measure-preserving) \textit{dynamical
system}. The working hypothesis of nonlinear time series analysis (NTSA) is
that real world (univariate) time series are scalar observations of a
generally higher dimensional dynamical system $(\Omega ,\mathcal{B},\mu ,f)$
which, moreover, is either unknown, only partially known, or too complex for
its knowledge to be of any practical use. This being the case, it is assumed
hereafter that any time series $(x_{n})_{n\geq 0}$, $x_{n}\in \mathbb{R}$,
is not, in general, an orbit of an autonomous dynamic $f$ but an orbit of a
non-autonomous dynamic $\varphi \circ f$, where $\varphi \colon \Omega
\rightarrow \mathbb{R}$ is a so-called \textit{observation function},
typically a projection $\mathbf{x}\mapsto x$ from $\Omega $ onto one of its
local coordinates. From the point of view of time series analysis, this
results in that the observations $(x_{n})_{n\geq 0}$ can be viewed as
random, unless $\Omega \subset \mathbb{R}$ and $\varphi $ is invertible
(e.g., the identity). For this reason, we will write 
\begin{equation*}
(x_{n})_{n\geq 0}=(\varphi \circ f^{n}(\mathbf{x}_{0}))_{n\geq 0}=(\varphi (%
\mathbf{x}_{n}))_{n\geq 0}=:\text{ }\mathbf{X}(\mathbf{x}_{0})
\end{equation*}%
and say that $(x_{n})_{n\geq 0}$ is a trajectory or realization of the
random process $\mathbf{X}$. Here $f^{n}$ is the $n$th iterate of $f$ for $%
n\geq 1$, $f^{0}$ is the identity, and $\mathbf{x}_{n}=f^{n}(\mathbf{x}%
_{0})\in \Omega $. Formally, $\mathbf{X}(\mathbf{x}_{0})$ is a deterministic
orbit with dynamical noise.

Because of the applications of NTSA to real world observations, specially to
irregular time series, the underlying dynamical system $(\Omega ,\mathcal{B}%
,\mu ,f)$ is generally supposed to be chaotic, and $\Omega $ to be an
attractor. Some techniques of NTSA aim to reconstruct and characterize the
attractors of such systems from the observations $(x_{n})_{n\geq 0}$ using
time delay coordinates \cite{Takens1981,Sauer1991,Stark1999}. Such characterization include Lyapunov exponents, several
dimensions and dynamical entropies such as the Kolmogorov-Sinai entropy \cite%
{Kantz1997,Sprott2003}. However, since our goal in this paper is to classify
patients into two groups based on their heart rate and blood pressure, it
will suffice to characterize the structure and complexity of those
recordings. To this end, there are two main options: linear (statistical)
and nonlinear tools. As the former have been well investigated already, we
focus here on the latter. Specifically, we will calculate various kinds of
entropy and mutual information using ordinal patterns, which are the
building blocks of the so-called ordinal methodology. This is a convenient
way to convert continuous-valued time series into finite-valued (or
symbolic-valued) time series for the reasons that we mention in the
following section.

%%%%%%%%%%%%%%%%%%%%%%%%%%%%%%%%%%%%%%%%%%%%%%%%%%%%%

%\FloatBarrier

\subsection{Ordinal patterns and transcripts}

\label{sec:ord}

To characterize the complexity of the time series $\mathbf{X}(\mathbf{x}%
_{0}) $ we will use the ordinal methodology, which trades off blocks $%
x_{n}^{L}:=x_{n},x_{n+1},...,x_{n+L-1}$ of length $L\geq 2$ for rank vectors
called ordinal patterns or simply permutations of length $L$ \cite{Bandt2002}%
. Ordinal patterns are conceptually simple and their calculation can be done
in real time because knowledge of the data range is not required in advance.
In addition, since ordinal patterns can be identified with permutations, one
can take advantage of their algebraic structure, e.g., through transcripts,
as we will do in Section \ref{sec:Transcripts}. Last but not least, the
Shannon entropy rate of the ordinal patterns as $L\rightarrow \infty $
equals the Kolmogorov-Sinai entropy of the underlying dynamics in the case
of one-dimensional interval maps under very mild assumptions \cite%
{Bandt2002B}, which shows that ordinal patterns can capture the complexity
of time series. For generalizations of the ordinal methodology to higher
dimensional maps, the interested reader is referred to \cite{Amigo2013}.

Next we introduce the conceptual and notational setting of the ordinal
methodology. Let $(r_{0},r_{1},...,r_{L-1})$ be a permutation of $%
\{0,1,...,L-1\}$, $L\geq 2$. We say that the block $x_{n}^{L}$ of $\mathbf{X}%
(\mathbf{x}_{0})$ defines the \textit{ordinal pattern of length} $L$ (or
simply ordinal $L$-pattern) $\mathbf{r}=(r_{0},r_{1},...,r_{L-1})$ if 
\begin{equation}
x_{n+r_{0}}<x_{n+r_{1}}<\ldots <x_{n+r_{L-1}}  \label{ord patt}
\end{equation}%
(other rules can be found in the literature). We use the notation $%
\mathbf{r}=\mathrm{rank}(x_{n}^{L})$ or $\mathbf{r}_{n}=\mathrm{rank}%
(x_{n}^{L})$ to associate ordinal patterns and blocks. Since $\mathbf{r}$ is
a permutation of $\{0,1,...,L-1\}$, sometimes one also speaks of
permutations instead of ordinal patterns and writes $\mathbf{r}\in \mathcal{S%
}_{L}$, where $\mathcal{S}_{L}$ denotes the set (actually a group) of such
permutations. Therefore, the number of ordinal $L$-patterns is $\left\vert 
\mathcal{S}_{L}\right\vert =L!$.

For example, if $L=5$ and 
\begin{equation}
x_{n}=0.4,\;x_{n+1}=-0.5,\;x_{n+2}=1.5,\;x_{n+3}=-0.8,\;x_{n+4}=1.4,\;
\label{Ex ord patt}
\end{equation}%
then $\mathrm{rank}(x_{n}^{5})=(3,1,0,4,2)$. In case of ties, several
conventions can be adopted. If the number of ties in $(x_{n})_{n\geq 0}$ is
small (as we will suppose here), it can be agreed that the earlier entry is
the smaller, or a small amplitude noise can be added to the signal to
eliminate the ties \cite{Myers2020}.

More generally, the \textit{ordinal representation} $\mathbf{R}(L)=(\mathbf{r%
}_{n})_{n\geq 0}$ of a time series $\mathbf{X}(\mathbf{x}_{0})=(x_{n})_{n%
\geq 0}$ is obtained by rank-ordering a sliding window of size $L$ along $%
\mathbf{X}(\mathbf{x}_{0})$, i.e., $\mathbf{r}_{n}=\mathrm{rank}(x_{n}^{L})$%
. Therefore, ordinal representations are finite-state random processes
because they are obtained by discretizing continuous-valued times series. In
practice, $\mathbf{X}(\mathbf{x}_{0})$ is also randomized by
observational noise.

Also in practice, time series have a finite length $N$. In this case, the
ordinal representation of $(x_{n})_{0\leq n\leq N-1}$ of parameter $L\leq N$
is $(\mathbf{r}_{n})_{0\leq n\leq N-L}$, a symbolic representation of length 
$N-L+1$. Furthermore, it may be convenient not to consider the complete
sequence $(x_{n})_{0\leq n\leq N-1}$ but only an equally spaced subsequence $%
(x_{m\tau })_{0\leq m\leq M-1}$, where $\tau \geq 1$ is called a \textit{%
delay time} (or lag) and $M=\left\lfloor \frac{N-1}{\tau }+1\right\rfloor $.
Therefore, an ordinal representation has in general two parameters: the
length $L\geq 2$ of the ordinal patterns (sometimes called the \textit{%
embedding dimension} of the representation) and the delay time $\tau $ ($=1$
unless otherwise stated). Delay times greater than 1 are used, for example,
when the sampling frequency of an analog signal is high compared to the
inverse of the typical time constant of the sampled signal. For methods to
select the parameters $L$ and $\tau $ and the subtleties involved, see e.g. 
\cite{Myers2020}.

%%%%%%%%%%%%%%%%%%%%%%%%%%%%%%%%%%%%%%%%%%%%%%%%%%%%%

%\FloatBarrier

\subsection{Entropies and entropy-like quantities based on ordinal patterns}

\label{sec:entropy}

From the toolbox of Information Theory, we will resort to the perhaps
simplest, uni- and two-variate quantities, namely, entropy and mutual
information.

%%%%%%%%%%%%%%%%%%%%%%%%%%%%%%%%%%%%%%%%%%%%%%%%%%%%%

%\FloatBarrier

\subsubsection{Permutation entropy}

\label{sec:perm}

Suppose that $\mathbf{X}(\mathbf{x}_{0})=(x_{n})_{n\geq 0}$ are scalar
observations of a chaotic dynamical system $(\Omega ,\mathcal{B},\mu ,f)$,
where $\Omega $ is a minimal attractor and, hence, $f$ is ergodic on $\Omega 
$. Then the probabilities $p(\mathbf{r})$ of the ordinal $L$-patterns
generated by $\mathbf{X}$ (i.e., the probability that \textrm{rank}$%
(x_{n}^{L})=\mathbf{r}$ for any $n\geq 0$) can be estimated by the 
empirical probabilities $\hat{p}_{N}(\mathbf{r})$, that is, the
normalized count of sliding windows $x_{n}^{L}$ with ordinal pattern $%
\mathbf{r}\in \mathcal{S}_{L}$. For a time
series of length $N\gg L$ we have 
\begin{equation}
\hat{p}_{N}(\mathbf{r})=\frac{\#\{n:\mathrm{rank}(x_{n}^{L})=\mathbf{r}%
,\,0\leq n\leq N-L\}}{N-L+1}.  \label{p_N(r)}
\end{equation}

The \textit{permutation entropy of order} $L$ of a process $\mathbf{X}$,
whether random or deterministic, is the Shannon entropy of the probability
distribution of ordinal $L$-patterns, $\{p(\mathbf{r}):\mathbf{r}\in 
\mathcal{S}_{L}\}$, generated by $\mathbf{X}$, i.e., 
\begin{equation}
\text{\texttt{X\_PE(L})}=-\sum_{\mathbf{r}\in \mathcal{S}_{L}}p(\mathbf{r}%
)\ln p(\mathbf{r}).  \label{PermEntr}
\end{equation}%
By convention, $0\cdot \ln 0:=\lim_{x\rightarrow 0+}x\ln x=0$. We use
natural logarithms (entropy in nats) throughout.

The maximum of the Shannon entropy is achieved for flat distributions \cite%
{Cover2006}. Therefore, the maximum of the permutation entropy of order $L$
occurs when $p(\mathbf{r})=1/L!$ for all $\mathbf{r}\in \mathcal{S}_{L}$,
which implies $\text{\texttt{X\_PE(L})}\leq \ln L!$. This bound prompts
to define the \textit{normalized permutation entropy} of order $L$,%
\begin{equation}
\text{\texttt{X\_NPE(L})}=\frac{\text{\texttt{X\_PE(L})}}{\ln L!}=-\frac{1}{%
\ln L!}\sum_{\mathbf{r}\in \mathcal{S}_{L}}p(\mathbf{r})\ln p(\mathbf{r}),
\label{NPE(L)}
\end{equation}%
so that%
\begin{equation}
0\leq \text{\texttt{X\_NPE(L})}\leq 1.  \label{NPE(L)2}
\end{equation}

%%%%%%%%%%%%%%%%%%%%%%%%%%%%%%%%%%%%%%%%%%%%%%%%%%%%%

%\FloatBarrier

\subsubsection{Statistical complexity}

\label{sec:Statcomplexity}

Among the entropy-like quantities that we will use in Section \ref{sec:appl}
for classification is the statistical (permutation) complexity of order $L$
of a continuous-valued process $\mathbf{X}$. Its definition, adapted to our
setting and notation, comprises two ingredients: the normalized permutation
entropy \text{\texttt{X\_NPE(L})}, Equation (\ref{NPE(L)}), and the \textit{%
desequilibrium} $Q(L)$, which we define next.

Let 
\begin{equation}
D_{JS}(L)=\text{\texttt{U\_PE(L})}-\frac{1}{2}\text{\texttt{X\_PE(L})}-\frac{%
1}{2}\ln L!  \label{SC1}
\end{equation}%
be the \textit{Jensen-Shannon divergence} from the probability distribution $%
\{p(\mathbf{r}):\mathbf{r}\in \mathcal{S}_{L}\}$ of the ordinal $L$-patterns
of the process $\mathbf{X}$ to the probability distribution 
\begin{equation}
\{p(\mathbf{u}):\mathbf{u}\in \mathcal{S}_{L}\}=\left\{ \frac{1}{2}\left( p(%
\mathbf{r})+\frac{1}{L!}\right) :\mathbf{r}\in \mathcal{S}_{L}\right\}
\label{SC2}
\end{equation}%
of the ordinal $L$-patterns of an auxiliary process $\mathbf{U}$. $D_{JS}(L)$
is a symmetrized version of the Kullback-Leibler divergence \cite{Cover2006}%
; its square root is a distance, in our case between the probability
distributions $\{p(\mathbf{r}):\mathbf{r}\in \mathcal{S}_{L}\}$ and $\{p(%
\mathbf{u}):\mathbf{u}\in \mathcal{S}_{L}\}$. Then \cite{Martinez2022} 
\begin{equation}
Q(L)=Q_{0}D_{JS}(L)  \label{SC3}
\end{equation}%
is the \textit{desequilibrium of order} $L$ of $\mathbf{X}$, where $Q_{0}$
is the normalization constant%
\begin{equation}
Q_{0}=-2\left[ \frac{L!+1}{L!}\ln (L!+1)-2\ln (2L!)+\ln L!\right] ^{-1}.
\label{SC4}
\end{equation}

Finally, the \textit{statistical complexity of order} $L$ of $\mathbf{X}$ is
defined as 
\begin{equation}
\text{\texttt{X\_SC(L})}=Q(L)\cdot\left(\text{\texttt{X\_NPE(L})}\right) .
\label{SC5}
\end{equation}

%%%%%%%%%%%%%%%%%%%%%%%%%%%%%%%%%%%%%%%%%%%%%%%%%%%%%

%\FloatBarrier

\subsection{Entropies and entropy-like quantities based on transcripts}

\label{sec:Transcripts}

As mentioned in Section \ref{sec:ord}, ordinal representations have an
additional property that can also be exploited in time series analysis,
namely:\ their symbols (permutations) belong to an algebraic group. Indeed, $%
\mathcal{S}_{L}$ is a group under the composition (or \textquotedblleft
product\textquotedblright ) of permutations, called the symmetric group of
degree $L$. Precisely, transcripts is perhaps the simplest way to capitalize
on the algebraic structure of $\mathcal{S}_{L}$.

Thus, consider two ordinal representations $(\mathbf{r}_{n})_{n\geq 0}$ and $%
(\mathbf{s}_{n})_{n\geq 0}$ of the time series $\mathbf{X}(\mathbf{x}%
_{0})=(x_{n})_{n\geq 0}$ and $\mathbf{Y}(\mathbf{y}_{0})=(y_{n})_{n\geq 0}$,
respectively, with the same embedding dimension $L\geq 2$. We say that the
symbolic sequence $(\mathbf{t}_{n}^{\Lambda })_{n\geq 0}$, where $\Lambda
\in \mathbb{Z}$ and $\mathbf{t}_{n}^{\Lambda }\in \mathcal{S}_{L}$ for all $%
n\geq 0$, is a transcript representation from the process $\mathbf{X}(%
\mathbf{x}_{0})$ to the process $\mathbf{Y}(\mathbf{y}_{0})$ with \textit{%
coupling delay} $\Lambda $ if%
\begin{equation}
\mathbf{t}_{n}^{\Lambda }=\mathbf{s}_{n+\Lambda }\circ \mathbf{r}_{n}^{-1},
\label{transcripts}
\end{equation}%
where $\mathbf{r}_{n}^{-1}$\textbf{\ }is the inverse of $\mathbf{r}_{n}$ in
the group $\mathcal{S}_{L}$ \cite{Monetti2009,Amigo2012}. Note that $%
\mathcal{S}_{L}$ is non-commutative for $L\geq 3$, so the order of the
factors in (\ref{transcripts}) matters. To distinguish the particular case $%
\mathbf{X}(\mathbf{x}_{0})=\mathbf{Y}(\mathbf{y}_{0})$ from the general case 
$\mathbf{X}(\mathbf{x}_{0})\neq \mathbf{Y}(\mathbf{y}_{0})$, one speaks of
self-transcripts and cross-transcripts, respectively.

The algebraic properties of transcripts and their generalization to more
than two time series have been studied in \cite{Monetti2013,Amigo2014}.

\subsubsection{Transcript entropy}

The \textit{transcript entropy of order} $L$ \textit{and} \textit{coupling
delay} $\Lambda $ from the process $\mathbf{X}$ to the process $\mathbf{Y}$
is the Shannon entropy of the time series $(\mathbf{t}_{n}^{\Lambda
})_{n\geq 0}$ (see Equation (\ref{transcripts})), i.e.,%
\begin{equation}
\text{\texttt{XY\_TE(L,}}\Lambda \text{)}=-\sum_{\mathbf{t}\in \mathcal{S}%
_{L}}p(\mathbf{t}^{\Lambda })\ln p(\mathbf{t}^{\Lambda })=\text{\texttt{XY}}%
^{\Lambda }\text{\texttt{\_TE(L}),}  \label{cross-transcript ent}
\end{equation}%
where $\mathbf{Y}^{\Lambda }\mathbf{=}(y_{n+\Lambda })_{n\geq \lbrack
-\Lambda ]_{+}}$ (i.e., $n\geq 0$ if $\Lambda \geq 0$ and $n\geq -\Lambda $
if $\Lambda <0$). Here the probabilities $p(\mathbf{t}^{\Lambda })$ are
estimated by the frequencies 
\begin{equation*}
\hat{p}_{N}(\mathbf{t}^{\Lambda })=\frac{\#\{n:\mathrm{rank}(x_{n}^{L})=%
\mathbf{r}\,\,\&\,\,\mathrm{rank}(y_{n+\Lambda }^{L})=\mathbf{t}^{\Lambda }%
\mathbf{\circ r},\,\,[-\Lambda ]_{+}\leq n\leq N-L\}}{N-L+1}
\end{equation*}%
for any $\mathbf{r}\in \mathcal{S}_{L}$, since then and only then $\mathrm{%
rank}(y_{n+\Lambda }^{L})\circ $ $(\mathrm{rank}(x_{n}^{L}))^{-1}=\mathbf{t}%
^{\Lambda }\mathbf{\circ r\circ r}^{-1}=\mathbf{t}^{\Lambda }$. Moreover, $N\gg L$ and $[-\Lambda ]_{+}\ll $ $N-L$ for statistical significance.

If $\mathbf{Y}(\mathbf{y}_{0})=\mathbf{X}(\mathbf{x}_{0})$ and $\Lambda \neq
0$, then we speak of self-transcript entropy,%
\begin{equation}
\text{\texttt{X\_STE(L,}}\Lambda \text{),}  \label{self-transcript ent}
\end{equation}%
to distinguish this special case from the general \textquotedblleft
cross-transcript\textquotedblright\ entropy (\ref{cross-transcript ent})
with $\mathbf{Y}(\mathbf{y}_{0})\neq \mathbf{X}(\mathbf{x}_{0})$. If the
data $\mathbf{X}(\mathbf{x}_{0})$ and $\mathbf{Y}(\mathbf{y}_{0})$ are clear
from the context, we shorten the notations (\ref{cross-transcript ent}) and (%
\ref{self-transcript ent}) to \texttt{TE(L,}$\Lambda $) and \texttt{STE(L,}$%
\Lambda $), respectively. If $\Lambda =0$, the parameter $\Lambda $ is
dropped.

%%%%%%%%%%%%%%%%%%%%%%%%%%%%%%%%%%%%%%%%%%%%%%%%%%%%%

%\FloatBarrier

\subsubsection{Permutation and transcript mutual information}

\label{sec:mutual}

The \textit{permutation mutual information} of order $L$ \textit{and
coupling delay} $\Lambda $ between the processes $\mathbf{X}$ and $\mathbf{Y}
$ is defined as 
\begin{equation}
\text{\texttt{XY\_MI(L,}}\Lambda \text{)}=\text{\texttt{X\_PE(L})}+\text{%
\texttt{Y\_PE(L})}-\text{\texttt{XY}}^{\Lambda }\text{\texttt{\_PE(L})}\geq
0,  \label{Perm Mutual Info}
\end{equation}%
where \texttt{XY}$^{\Lambda }$\texttt{\_PE(L}) is the permutation entropy of
order $L$ of the joint probability distribution $\{p(\mathbf{r},\mathbf{s}%
^{\Lambda }):\mathbf{r,s}^{\Lambda }\mathbf{\in }\mathcal{S}_{L}\}$, where $%
p(\mathbf{r},\mathbf{s}^{\Lambda })$ is the probability that \textrm{rank}$%
(x_{n}^{L})=\mathbf{r}$ and \textrm{rank}$(y_{n+\Lambda }^{L})=\mathbf{s}%
^{\Lambda }$ for any $n\geq 0$.

Finally, we will also use a particular sort of transcript-based mutual
information that was proposed in \cite{Monetti2013B} as a causality
indicator. We define the \textit{transcript mutual information of order }$L$ 
\textit{and coupling delay} $\Lambda $ from $\mathbf{X}$ to $\mathbf{Y}$ as%
\begin{equation}
\text{\texttt{XY\_TMI(L,}}\Lambda \text{)}=\text{\texttt{XY\_TE(L})}+\text{%
\texttt{Y}}^{\Lambda }\text{\texttt{Y\_TE(L})}-\text{\texttt{XY\_Y}}%
^{\Lambda }\text{\texttt{Y\_TE(L}),}  \label{TranscriptMI}
\end{equation}%
where (see Equation (\ref{cross-transcript ent})) \texttt{XY\_TE(L)}, 
\texttt{Y}$^{\Lambda }$\texttt{Y\_TE(L)} and\texttt{\ XY\_Y}$^{\Lambda }$%
\texttt{Y\_TE(L}) are the Shannon entropies of the transcript probability
distributions $\{p(\mathbf{s}\circ \mathbf{r}^{-1}):\mathbf{r},\mathbf{s}\in 
\mathcal{S}_{L}\}$, $\{p(\mathbf{s}\circ (\mathbf{s}^{\Lambda })^{-1}):%
\mathbf{r},\mathbf{s}^{\Lambda }\in \mathcal{S}_{L}\}$, and the joint
transcript probability distribution $\{p(\mathbf{s}\circ \mathbf{r}^{-1},%
\mathbf{s}\circ (\mathbf{s}^{\Lambda })^{-1}):\mathbf{r},\mathbf{s},\mathbf{s%
}^{\Lambda }\in \mathcal{S}_{L}\}$, respectively.

While the permutation mutual information \texttt{XY\_MI(L,}$\Lambda $\texttt{%
)} is symmetric under the exchange of $\mathbf{X}$ and $\mathbf{Y}$ (hence,
useless for detecting the information direction in coupled processes), for
the transcript mutual information \texttt{XY\_TMI(L,}$\Lambda $\texttt{%
)\thinspace }$\neq $ \texttt{YX\_TMI(L,}$\Lambda $\texttt{)}. If the
processes $\mathbf{X}$ and $\mathbf{Y}$ are clear from the context, we
shorten the previous notation to \texttt{MI(L,}$\Lambda $) and \texttt{TMI(L,%
}$\Lambda $). The transcript mutual information is a two-dimensional
approximation of the three dimensional symbolic transfer entropy \cite%
{Staniek2008} and it was shown in \cite{Amigo2015,Amigo2016,Amigo2019} to be
an efficient indicator of information directionality.

%%%%%%%%%%%%%%%%%%%%%%%%%%%%%%%%%%%%%%%%%%

\end{document}